\documentclass[prd,showpacs,amsmath,amssymb]{revtex4-2}
\usepackage{amsmath}
\usepackage{amsfonts}
\usepackage{mathrsfs}
\usepackage{pstricks}
\usepackage{color}
\usepackage{graphicx}
\usepackage{slashed}
\usepackage{amssymb}
\usepackage{hyperref}

\newcommand{\be}{\begin{equation}}
\newcommand{\ee}{\end{equation}}
\newcommand{\ba}{\begin{array}{c}}
\newcommand{\ea}{\end{array}}
\newcommand{\bqa}{\begin{eqnarray}}
\newcommand{\eqa}{\end{eqnarray}}
\makeatletter
    
    \newcommand{\Rmnum}[1]{\expandafter\@slowromancap\romannumeral #1@}
    \makeatother

\begin{document}

\title{\bf On-shell renormalization in NREFT: a simplified description of
resonances, bound states, and coupled channel effects.}

\author{Guo-Ying Chen}

\affiliation{Department of Physics and Astronomy, Hubei University
of Education, Wuhan 430205, China}


\date{\today}

\begin{abstract}
We discuss several issues regarding the modern generalization of
Weinberg's compositeness criterion. In particular, we demonstrate
that the on-shell renormalization in NREFT reproduces Weinberg's
result and provides a simple description of resonances, bound
states, and coupled channel effects. Additionally, we present the
complete propagator, including the charged channel for the
near-threshold p-wave states.
\end{abstract}



\maketitle

\section{Introduction}

Nonrelativistic effective field (NREFT) for hadrons was first
proposed by
Weinberg~\cite{Weinberg:1990rz,Weinberg:1991um,Weinberg:1992yk} to
describe nucleon-nucleon interactions, and was further developed in
Refs.~\cite{Ordonez:1992xp,Ordonez:1993tn,vanKolck:1994yi,Ordonez:1995rz,Kaplan:1996nv,Kaplan:1996xu,Kaplan:1998tg,Kaplan:1998we}.
Over the past two decades, NREFT has been extensively applied to
meson-meson interactions following the discovery of many exotic
states (see Ref.~\cite{Dai:2026fkg} for a recent review). Current
debates on the nature of near-threshold exotic states center on
whether they are loosely bound molecular states or compact
multiquark states. This question relates to Weinberg's compositeness
criterion~\cite{Weinberg:1962hj,Weinberg:1963zza,Weinberg:1965zz},
which is formulated through relations among the binding energy B,
the field renormaliaztion constant $Z$, and the effective range
expansion parameters defined in the low-energy partial-wave
scattering amplitude

\begin{eqnarray}
\mathcal{A}_{\ell}&=&\frac{2\pi}{\mu}\frac{(2\ell
+1)p^{2\ell}}{p^{2\ell+1}
\cot\delta_\ell-ip^{2\ell+1}}P_\ell(\cos\theta),\nonumber\\
p^{2\ell+1}\cot \delta_\ell
&=&-\frac{1}{a_\ell}+\frac{r_\ell}{2}p^2-\frac{\mathcal{P}_\ell}{4}p^4+\cdots.\label{pwa}
\end{eqnarray}
Here $p$ is the center-of-mass momentum, $\ell=0,1,2,\cdots$ is the
quantum number of the angular momentum, and $\mu$ is the reduced
mass of the two-particle system. $a_\ell$, $r_\ell$,
$\mathcal{P}_\ell$ are the scattering length, effective range, and
shape parameter respectively. For the s-wave scattering, Weinberg
derived the following relations~\cite{Weinberg:1965zz}

\begin{equation}
a_0=\frac{2(1-Z)}{2-Z}\frac{1}{\sqrt{2\mu B}},\ \ \ \ \ \
r_0=-\frac{Z}{1-Z}\frac{1}{\sqrt{2\mu B}}.\label{ar}
\end{equation}

Modern generalization of Weinberg's criterion is to study the
compositeness, i.e., $X\equiv 1-Z$, of the exotic states. However,
it is found that both the compositeness $X$ and $Z$ are in general
complex for resonances. This motivates many theoretical discussions,
see a review paper~\cite{Kinugawa:2024crb} and references therein.

The aim of this manuscript is to clarify several issues regarding
the modern generalization of Weinberg's compositeness criterion from
the viewpoint of NREFT. Our paper is organized as follows. In
Sec.~II we discuss some aspects of NREFT. In Sec.~III we treat bound
states and resonances in the NREFT framework. Coupled channel
effects are addressed in Sec.~IV. We then summarize our results in
Sec.~V.

\section{Some aspects of NREFT}

Weinberg's compositeness criterion is incorporated into NREFT in
Refs.~\cite{Chen:2013upa,Xu:2024vne,Chen:2026zpl}. In particular,
Ref.~\cite{Chen:2026zpl} found that Weinberg's composite relations are equivalent to the on-shell renormalization scheme
in NREFT: (i) the full propagator of the near-threshold state
has a bound state pole. Note that in our language a bound state is a
below-threshold state, which should be distinguished from a
molecular state (not only below threshold but also with $Z=0$); (ii)
the residue of the propagator at this pole is $Z$. This allows us to
obtain the compositeness relations in Eq.\eqref{ar} without invoking
the normalization condition of the bound state wave function, which
was used in Weinberg's original derivation. It is then
straightforward to generalize to the p-wave near-threshold bound
states, and the corresponding compositeness relations for p-wave read~\cite{Chen:2026zpl}

\begin{equation}
a_1=\frac{1-Z}{\Lambda-(2+Z)\Lambda_0/3}\frac{1}{2\mu B},\ \ \
r_1=-\frac{2}{1-Z}(\Lambda-Z\Lambda_0),\label{CRP}
\end{equation}
where $\Lambda$ is the renormalization scale and $\Lambda_0=\frac{3}{2}\sqrt{2\mu B}$.

One advantage of NREFT is that it makes use the systematic expansion
of $p/Q$, where $p$ is the typical momentum scale of the low energy effective field theory, and $Q$ is the energy scale characterized by the
heavier degrees of freedom which are not included explicitly in the effective field theory. We treat $p$, $\Lambda$ and
$\gamma=\sqrt{2\mu B}$ as being of the same order, i.e.,
$\mathcal{O}(p)$. One can find that the three terms
$-\frac{1}{a_\ell}$, $\frac{r_\ell}{2}p^2$ and $-ip^{2\ell+1}$ in
the denominator of $\mathcal{A}_\ell$ in Eq.\eqref{pwa} are of the
same order for s- and p-waves. This is an important observation,
because extensive literature states that the s-wave amplitude is
characterized only by the scattering length for small $p$, see the
review paper~\cite{Kinugawa:2024crb} and references therein. This
feature is usually called low-energy universality. From
Eq.\eqref{ar}, one can see that the low-energy universality is
certainly valid for $Z=0$, which is exactly the case in the nucleon
systems. However, if $Z\neq 0$, $-\frac{1}{a_0}$ and
$\frac{r_0}{2}p^2$ are of the same order, i.e., $\mathcal{O}(p)$
in the power counting. Therefore, the low-energy universality cannot
be applied for $Z\neq 0$. In other words, for $Z\neq 0$, the
low-energy scattering amplitude has at least two-independent
parameters, either expressed as $(a_\ell, r_\ell)$ or ($B, Z$). This
makes it impossible, in principle, to prove $Z\to 0$ in the limit
$B\to 0$. For example, although the famous $X(3872)$ has a very
small binding energy, it is widely accepted that it has a
non-negligible $Z$ value~\cite{Dai:2026fkg,Xu:2023lll}.

The other aspect of NREFT is that it has the general feature of
field theory, the emergence of the renormalization scale. This is
evident by adopting the power divergence subtraction (PDS)
scheme~\cite{Kaplan:1998we} in loop integrals. It is shown that the
leading-order s-wave scattering amplitude happens to exclude the
renormalization scale $\Lambda$
explicitly~\cite{Chen:2013upa}, but this scale must be explicitly
included and should be chosen as $\Lambda>\Lambda_0$ to avoid the negative-norm state in the p-wave sector~\cite{Chen:2026zpl}. This implies
that in general, $Z$ depends on the renormalization scale, as the
resulting amplitude must be renormalization-scale-independent. This
is not surprising, as it is just the standard result in the
textbooks of quantum field theory. It indicates that $Z$ may not
have a probability interpretation. We note that a similar point was
pointed out in Refs.~\cite{Nagahiro:2014mba,Esposito:2021vhu}. Thus
Weinberg's criterion may be understood as: the near-threshold state
is a pure molecular state for $Z=0$ and an elementary state for
$Z\neq 0$~\cite{Nagahiro:2014mba}.

\section{Bound states and resonances in NREFT}

We study the bound states and resonances in NREFT in this section.
For simplicity, we only discuss the s-wave coupling. Consider a bare
state $X$ with bare mass $-B_0$ and coupling $g_0$ to the
two-particle state $D\bar{D}$. Note that the bare mass is defined
relative to the two-particle threshold; thus, $B_0>0$ indicates that
a bare state lies below the threshold. The interaction Lagrangian
reads
\begin{equation}
\mathcal{L}_{\mathrm{int}}=g_0 XD\bar D.
\end{equation}

\begin{figure}[hbt]
 \begin{center}
  \includegraphics[width=10cm]{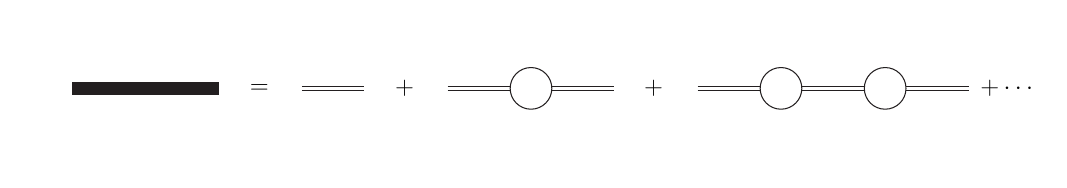}\\
  \caption{Full propagator for near-threshold states. The double line denotes the bare
  state.
  The particles in the loop are intermediate states.  }\label{ppgator}
  \end{center}
 \end{figure}

Hereafter, we adopt the convention in Ref. \cite{Chen:2013upa},
where the boson field operator includes a factor of $\sqrt{2M}$ so
that it has mass dimension $3/2$. The Feynman rule for the external
boson line is $\sqrt{2M}$ with the convention. We first assume $X$
is an above-threshold resonance, i.e., $m_X=-B_0$ with $B_0<0$. The
decay amplitude for $X\to D\bar D$ is
\begin{equation}
i\mathcal{M}=ig_0\sqrt{2m_X}\cdot\sqrt{2m_1}\cdot\sqrt{2m_2},
\end{equation}
where $m_1,m_2$ are the masses of $D,\bar D$ respectively. The decay
width for $X\to D\bar D$ then reads
\begin{equation}
\Gamma_0=g_0^2\frac{p}{\pi}\frac{m_1 m_2}{m_X},
\end{equation}
where $p$ is the momentum of $D$ in the center-of-mass frame. If
$m_X$ is near the threshold, we can have $m_X\approx m_1+m_2$ then
\begin{equation}
\Gamma_0=g_0^2\frac{\mu}{\pi}p.\label{gamma0}
\end{equation}
We then come to study the propagator of $X$. Near the two-particle
threshold, and in the center-of-mass frame, the full propagator of
$X$ in Fig.\ref{ppgator} takes the form
\begin{equation}
i\Delta_X=\dfrac{i}{E+B_0+\Sigma(E)},\label{deltaX}
\end{equation}
where the energy $E$ is defined relative to the $D\bar D$ threshold,
and $\Sigma(E)$ is the self-energy function. In NREFT with the PDS scheme, it reads 
\begin{equation}
\Sigma(E)=-g_0^2\frac{\mu}{2\pi}(\sqrt{-2\mu
E-i\epsilon}-\Lambda).\label{sigmaE}
\end{equation}
Using $\sqrt{-2\mu E-i\epsilon}=-ip$, we have
\begin{equation}
\Sigma(E)=ig_0^2\frac{\mu}{2\pi}p+g_0^2\frac{\mu\Lambda}{2\pi}.\label{RsigmaE}
\end{equation}
Comparing Eq.\eqref{gamma0} with Eq.\eqref{RsigmaE}, we can find
that
\begin{equation}
\Sigma(E)=i\Gamma_0/2+\tilde{B}_0-B_0, \label{rfsingma}
\end{equation}
with $\tilde{B}_0=B_0+g_0^2\frac{\mu\Lambda}{2\pi}$.
Then the full propagator of $X$ can be
rewritten as
\begin{equation}
i\Delta_X=\frac{i}{E+\tilde{B}_0+i\Gamma_0/2}.
\end{equation}
This is the near-threshold approximation of the Breit-Wigner
propagator~\cite{Xu:2024vne}, which is used for unstable resonances.
We now adopt the
conventional on-shell (OS) renormalization scheme: (i) the physical
mass is defined as the energy E at which the real part of the
denominator of the propagator, i.e., Eq.\eqref{deltaX}, vanishes, or equivalently,
\begin{equation}
m_X=-B_0-\mathrm{Re}\ \Sigma(m_X).\label{os1}
\end{equation}
(ii) The field renormalization constant $Z$ is defined as the
residue of the propagator at the physical pole, after the imaginary
part of the denominator is discarded (not by taking the real part,
but by simply omitting the imaginary part), or explicitly,
\begin{equation}
Z=\frac{1}{1+\mathrm{Re}\ \Sigma^\prime(m_X)}.\label{os2}
\end{equation}
With the OS scheme, we can find that for an above-threshold
resonance, we have $Z=1$, as $\Sigma^\prime(m_X)$ is purely imaginary in
NREFT. We would like to emphasize that this conclusion holds
provided that $X$ couples only to channels with threshold lighter
than its mass, and does not couple to any heavier channels; this
will become evident as we proceed.

We now turn to the case where $X$ is a below $D\bar D$ threshold
bound state, i.e., $B_0>0$. In this case, $\Sigma(m_X)$ is purely
real, so the physical mass of $X$ is shifted from $-B_0$ to $-B$,
where $B>0$, is the binding energy. Then the OS scheme gives
\begin{equation}
B=B_0+\mathrm{Re}\ \Sigma(-B)=B_0-g_0^2\frac{\mu}{2\pi}(\sqrt{2\mu
B}-\Lambda).\label{BB0}
\end{equation}
Since $Z$ is the residue of $\Delta_X$ at the bound state pole, we
have
\begin{equation}
g_0^2=\frac{2\pi\sqrt{2\mu B}}{\mu^2}\frac{1-Z}{Z},\label{g02}
\end{equation}
which is just the normalization condition of the bound state wave function.
Substituting Eq.\eqref{g02} into Eq.\eqref{BB0} gives
\begin{equation}
B_0=\frac{2-Z}{Z}B-\frac{1-Z}{Z}\sqrt{2B/\mu}\Lambda.\label{B0B}
\end{equation}
Substituting Eq.\eqref{g02} and Eq.\eqref{B0B} into the full
propagator Eq.\eqref{deltaX} gives
\begin{eqnarray}
i\Delta_X&=&\frac{iZ}{E+B+\tilde{\Sigma}(E)},\nonumber\\
\tilde{\Sigma}(E)&=&-g^2[\frac{\mu}{2\pi}\sqrt{-2\mu{E}-i\epsilon}+\frac{\mu\sqrt{2\mu{B}}}{4\pi{B}}(E-B)],\nonumber\\
g^2&=&Z g_0^2=\frac{2\pi\sqrt{2\mu B}}{\mu^2}(1-Z).
\end{eqnarray}
This is the general s-wave propagator given in
Refs.~\cite{Chen:2013upa,Xu:2024vne} after taking the limit
$\Gamma\to 0$ there. This propagator is
equivalent to Weinberg's compositeness relations, i.e,
Eq.\eqref{ar}, see Refs.~\cite{Xu:2024vne,Chen:2026zpl} for details.
One can see that while the field renormalization constant for the
above-threshold resonance is unity, the corresponding constant for
the below-threshold bound state can be different from unity; it
ranges from zero to unity from the K\"all\'en-Lehmann
representation.

\section{Coupled-channel effects in NREFT}

We have only considered a single channel in Sec.~III; we now extend
the discussion to coupled channels in this section. Suppose the bare
state $X$ couples to two different channels, $C\bar C$ and $D\bar
D$, and the interaction Lagrangian is given by
\begin{equation}
\mathcal{L}_{\mathrm{int}}=g_0^\prime X C\bar C+g_0 X D\bar D.
\end{equation}
We assume that the $D\bar D$ threshold $M_{D\bar D}$ is higher than
the $C\bar C$ threshold $M_{C\bar C}$, and that the bare mass of
$X$, i.e., $M_0$, lies between them, or equivalently, $M_{C\bar
C}<M_0<M_{D\bar D}$. For convenience, we redefine the bare mass of
$X$ as $-B_0$, which is defined relative to the $D\bar D$ threshold.
Since $M_0$ lies between the two thresholds, $B_0$ is positive. The
full propagator of $X$ reads
\begin{equation}
i\Delta_X^C=\dfrac{i}{E+B_0+\Sigma_C(E)+\Sigma_D(E)},\label{ccdeltaX}
\end{equation}
where $E$ is also defined relative to the $D\bar D$ threshold,
$\Sigma_C(E)$ is the self-energy function from the $C\bar C$ loop,
and $\Sigma_D(E)$ is the corresponding function from the $D\bar D$
loop. As discussed in Sec.~III, we can replace the term
$\Sigma_C(E)$ with Eq.\eqref{rfsingma}, since $M_0>M_{C\bar C}$. Then the
coupled-channel propagator reads
\begin{equation}
i\Delta_X^C=\dfrac{i}{E+\tilde{B}_0+\Sigma(E)+i\Gamma_0/2},\label{cc2deltaX}
\end{equation}
where $\Sigma(E)$ is defined in Eq.\eqref{sigmaE} with $\mu$ the
reduced mass of $D\bar D$. The coupled-channel propagator differs
from the single-channel one for bound state by an additional pure
imaginary term $i\Gamma_0$. Since the pure imaginary term in the
self-energy function does not affect the definitions of the physical
mass and the field renormalization constant in the OS scheme, the
relations in Eqs.\eqref{BB0},\eqref{g02},\eqref{B0B} still hold.
Substituting Eqs.\eqref{g02},\eqref{B0B} (after simply replacing $B_0$ with $\tilde{B}_0$ there) into Eq.\eqref{cc2deltaX},
we straightforwardly obtain the general propagator proposed in
Ref.~\cite{Chen:2013upa,Xu:2024vne} (denoted by $G_X(E)$ there),

\begin{eqnarray}
i\Delta_X^C(E)&=&\frac{iZ}{E+B+\tilde{\Sigma}(E)+i\Gamma/2},\nonumber\\
\tilde{\Sigma}(E)&=&-g^2[\frac{\mu}{2\pi}\sqrt{-2\mu{E}-i\epsilon}+\frac{\mu\sqrt{2\mu{B}}}{4\pi{B}}(E-B)],\nonumber\\
g^2&=&Z g_0^2=\frac{2\pi\sqrt{2\mu B}}{\mu^2}(1-Z),\ \ \
\Gamma=Z\Gamma_0.\label{GX}
\end{eqnarray}

It may be necessary to further include an additional channel
$D_c\bar D_c$ with the threshold $M_{D_c\bar D_c}$ slightly higher
than the $D\bar D$ threshold. For example, the famous $X(3872)$, as
an isospin singlet state, couples both the neutral channel $D\bar D$
and the charged channel $D_c\bar D_c$. Hence the interaction
Lagrangian can be extended to read
\begin{equation}
\mathcal{L}_{\mathrm{int}}=g_0^\prime X C\bar C+g_0 X D\bar
D+g_{0c}X D_c\bar D_c.
\end{equation}
Then the full propagator reads
\begin{eqnarray}
i\Delta_X^{cc}&=&\frac{i}{E+\tilde{B}_0+\Sigma_D(E)+\Sigma_{D_c}(E)+i\Gamma_0/2},\nonumber\\
\Sigma_D(E)&=&-g_0^2\frac{\mu}{2\pi}(\sqrt{-2\mu
E-i\epsilon}-\Lambda),\nonumber\\
\Sigma_{D_c}(E)&=&-g_{0c}^2\frac{\mu_c}{2\pi}(\sqrt{-2\mu_c
(E-\delta)-i\epsilon}-\Lambda),\label{3872bare}
\end{eqnarray}
where $\mu_c$ is the reduced mass of $D_c\bar D_c$. $\delta$ is the
mass splitting between the charged channel and neutral channel, and
is included since $E$ is defined relative to the neutral channel.
The OS scheme (or Eqs.\eqref{os1},\eqref{os2}) gives
\begin{eqnarray}
\tilde{B}_0&=&B+g_0^2\frac{\mu}{2\pi}(\sqrt{2\mu
B}-\Lambda)+g_{0c}^2\frac{\mu_c}{2\pi}(\sqrt{2\mu_c(B+\delta)}-\Lambda),\nonumber\\
1&=&Z+Zg_0^2\frac{\mu^2}{2\pi\sqrt{2\mu
B}}+Zg_{0c}^2\frac{\mu_c^2}{2\pi\sqrt{2\mu_c(B+\delta)}}.\label{oseq}
\end{eqnarray}
These relations were first given in Ref.~\cite{Xu:2024vne} via a
more complicated approach. We now show that they are easy to obtain
with the OS scheme. Using Eq.\eqref{oseq}, we can reformulate
Eq.\eqref{3872bare} as
\begin{eqnarray}
i\Delta_X^{cc}&=&\frac{iZ}{E+B+\widetilde{\Sigma}_{cc}(E)+i\Gamma/2},\nonumber\\
\widetilde{\Sigma}_{cc}(E)&=&-g^2[\frac{\mu}{2\pi}\sqrt{-2\mu{E}-i\epsilon}+\frac{\mu\sqrt{2\mu
B}}{4\pi B}(E-B)]\nonumber\\
&&-g_c^2[\frac{\mu_c}{2\pi}\sqrt{-2\mu_c(E-\delta)-i\epsilon}+\frac{\mu_c\sqrt{2\mu_c(B+\delta)}}{4\pi(B+\delta)}(E-B-2\delta)],\label{R3872}
\end{eqnarray}
where $\Gamma=Z\Gamma_0$, $g^2=Zg_0^2$ and $g_c^2=Z g_{0c}^2$. Thus the second equation in Eq.\eqref{oseq} can be written as
\begin{equation}
g^2\frac{\mu^2}{2\pi\sqrt{2\mu
B}}+g_{c}^2\frac{\mu_c^2}{2\pi\sqrt{2\mu_c(B+\delta)}}=1-Z.
\end{equation}
Eq.\eqref{R3872} was first given in Ref.~\cite{Xu:2024vne}; we
repeat it for the completeness of our discussions. This above study
can be extended to the near-threshold p-wave state
straightforwardly. Consider a spin-one particle $\psi$ that has the mass $-B_0$ and couples to the p-wave $D\bar {D}$ and $D_c\bar
D_c$ channels. Here $D\bar D$ denotes the neutral channel and
$D_c\bar D_c$ denotes the charged channel. Again, the mass
$-B_0$ is defined relative to the $D\bar D$ threshold. We assume the
 mass of $\psi$ is below the $D\bar D$ threshold thus $B_0>0$.
The interaction Lagrangian reads
\begin{equation}
\mathcal{L}_{\mathrm{int}}=ig_{p0}\{D^\dagger\bm{\nabla} \bar
D-\bm{\nabla} D^\dagger\bar
D\}\cdot\bm{\psi}+ig_{pc0}\{D_c^\dagger\bm{\nabla} \bar
D_c-\bm{\nabla} D_c^\dagger\bar D_c\}\cdot\bm{\psi}+h.c.,
\end{equation}
Then the full propagator reads
\begin{eqnarray}
G(E)&=&i\delta^{ij}\Delta_G(E), \nonumber\\
\Delta_G(E)&=&\frac{1}{E+B_0+\Sigma_p(E)+i\Gamma_0/2},\nonumber\\
\Sigma_p(E)&=&-\frac{2\mu}{3\pi}g_{p0}^2(2\mu E) [(-2\mu
E-i\epsilon)^{1/2}-\Lambda]-\frac{2\mu_c}{3\pi} g_{pc0}^2[2\mu_c(E-\delta)]\{[-2\mu_c
(E-\delta)-i\epsilon]^{1/2}-\Lambda\}, \label{PPGp}
\end{eqnarray}
where $E$ is defined relative to the $D\bar D$ threshold, and $\Gamma_0$ is to account for the
decay of $\psi$ into channels with thresholds lower than the mass of
$\psi$. With the OS scheme, we have
\begin{eqnarray}
B_0&=&B-\frac{2\mu}{3\pi} g_{p0}^2(2\mu B
)[(2\mu
B)^{1/2}-\Lambda]-\frac{2\mu_c}{3\pi}
g_{pc0}^2[2\mu_c(B+\delta)]\{[2\mu_c(B+\delta)]^{1/2}-\Lambda\},\nonumber\\
1&=&Z+\frac{4\mu^2}{3\pi} Zg_{p0}^2(\Lambda-\Lambda_0)+\frac{4\mu_c^2}{3\pi}
Zg_{pc0}^2(\Lambda-\Lambda_{0c}),\label{osp}
\end{eqnarray}
with $\Lambda_0=\frac{3}{2}\sqrt{2\mu B}$ and $\Lambda_{0c}=\frac{3}{2}\sqrt{2\mu_c (B+\delta)}$.
Substituting Eq.\eqref{osp} into Eq.\eqref{PPGp}, we can reformulate
the full propagator as
\begin{eqnarray}
G(E)&=&i\delta^{ij}\Delta_G(E),\nonumber\\
\Delta_G(E)&=&\frac{Z}{E+B+\tilde{\Sigma}_p(E)+i\Gamma/2},\nonumber\\
\tilde{\Sigma}_p(E)&=& g_p^2\frac{2\mu}{3\pi}[(-2\mu
E-i\epsilon)^{3/2}+\mu\sqrt{2\mu B}(3E+B)]\nonumber\\
&&+
g_{pc}^2\frac{2\mu_c}{3\pi}\{[-2\mu_c(E-\delta)-i\epsilon]^{3/2}+\mu_c\sqrt{2\mu_c(B+\delta)}(3E+B-2\delta)\},
\end{eqnarray}
where $\Gamma=Z\Gamma_0$. $g_p^2=Z g_{p0}^2$, $g_{pc}^2=Z g_{pc0}^2$
and they satisfy
\begin{equation}
Z+\frac{4\mu^2}{3\pi}g_{p}^2(\Lambda-\Lambda_0)+\frac{4\mu_c^2}{3\pi}
g_{pc}^2(\Lambda-\Lambda_{0c})=1,\label{gpgpceq}
\end{equation}
which can be obtained from the second equation of Eq.\eqref{osp}. As the above equation is equivalent to the normalization condition of the bound state wave function, to ensure that the contributions from both the neutral and charged channels are strictly positive-definite, the renormalization scale should be chosen as $\Lambda>\Lambda_{0c}$. Since $\Lambda_{0c}>\Lambda_0$,  under this condition, no individual channel can generate a negative-norm contribution.
Setting $g_{pc}=0$, we recover the result of
Ref.~\cite{Chen:2026zpl} where the charged channel is not included.  Differentiate both sides of Eq.\eqref{gpgpceq} with respect to $\Lambda$, while noting that the renormalized coupling constants $g_p^2$ and $g_{pc}^2$ are independent of $\Lambda$, yields the renormalization group equation for $Z$. If we further assume exact isospin symmetry, i.e., $g_p^2=g_{pc}^2$, we obtain
\begin{equation}
\frac{dZ}{d\Lambda}=\frac{(\mu^2+\mu_c^2)(Z-1)}{\mu^2(\Lambda-\Lambda_0)+\mu_c^2(\Lambda-\Lambda_{0c})}.
\end{equation}
Its solution reads
\begin{equation}
Z=1-C(\Lambda-\Lambda_0^{\mathrm{eff}}),\ \ \ \mathrm{with}
\ \ \ \Lambda_0^{\mathrm{eff}}=\frac{\mu^2\Lambda_0+\mu_c^2\Lambda_{0c}}{\mu^2+\mu_c^2},\end{equation}
where $C$ is a positive number. Note that although the strict positivity of individual channels requires $\Lambda>\Lambda_{0c}$, the positivity of the combined solution only requires $\Lambda>\Lambda_0^{\mathrm{eff}}$, which is a weaker condition since $\Lambda_0^{\mathrm{eff}}<\Lambda_{0c}$.

Although $Z$ depends on $\Lambda$, Ref~\cite{Chen:2026zpl} argued that it can still be used to study the nature of the near-threshold states. The conclusion is: if $C$ is of order $\mathcal{O}(p^{-1})$, $Z$ will depend strongly to $\Lambda$ implying that the compact component in the bound state wave function is negligible. In contrast, if $C$ is of order $\mathcal{O}(Q^{-1})$, the $\Lambda$-dependence of $Z$ is weak, and $Z$ remains close to unity
for $\Lambda\sim \mathcal{O}(p)$. In this case, the near-threshold state can be identified as a compact state.

\section{Summary}

If a bound state exists, the low-energy two-body scattering
amplitude has in general two independent parameters, i.e., the
binding energy $B$ and the field renormalization constant $Z$. We
show how to formulate the propagator of the bound state with these
parameters in NREFT using the on-shell renormalization scheme. It is
then straightforward to generalize to resonances and to include
coupled channel effects, without making further assumptions or
performing analytical continuations. With this approach, we do not
encounter the complex $Z$ as mentioned in the literature. Although
$Z$ ranges from zero to unity, we argue that it may not simply have
a probability interpretation, as it is not a physical observable and
may depend on the renormalization scale. However, using the propagators presented in this manuscript to fit the lineshape data, one can obtain the value of $Z$. If $Z$ is close to zero, one can claim that it is a
pure molecular state, while if $Z\neq 0$, one should identify it as
a compact state.





%

\end{document}